\documentclass{article}
\usepackage[a4paper, textwidth=148mm]{geometry}
\usepackage[utf8]{inputenc}
\usepackage[T1]{fontenc}
\usepackage{graphicx}
\usepackage[version=3]{mhchem} % Formula subscripts using \ce{}
\usepackage{authblk}
\usepackage{biblatex}
\title{A multimodal \textit{operando}  neutron study of the phase evolution in a graphite electrode}

\author[1,2]{Monica-Elisabeta L\u{a}c\u{a}tu\c{s}u}
\author[2,*]{Luise Theil Kuhn}
\author[2]{Rune E. Johnsen}
\author[3,4]{Patrick K. M. Tung}
\author[5]{S\o ren Schmidt}
\author[6]{Takenao Shinohara}
\author[6]{Ryoji Kiyanagi}
\author[7]{Anton S. Tremsin}
\author[3,4]{Nancy Elewa}
\author[3,5]{Robin Woracek}
\author[2,3,8,*]{Markus Strobl}

\affil[1]{Technical University of Denmark, Department of Energy Conversion and Storage, Fysikvej, DK-2800, Kgs. Lyngby, Denmark}
\affil[2]{Laboratory for Neutron Scattering and Imaging, Paul Scherrer Institute, CH-5232, Villigen, Switzerland}
\affil[3]{Nuclear Physics Institute, ASCR, v. v. i. CZ - 250 68 $\check{R}$e$\check{z}$, Czech Republic}
\affil[4]{Institute of Physics, Na Slovance 1999/2, 182 00 Praha 8, Czech Republic}
\affil[5]{European Spallation Source, P.O. Box 176, S - 221 00 Lund, Sweden}
\affil[6]{J-PARC Center, Japan Atomic Energy Agency, Tokai, 319-1195, Japan}
\affil[7]{Space Sciences Laboratory, University of California at Berkeley, Berkeley, CA, 94720, USA}
\affil[8]{Niels Bohr Institute, University of Copenhagen, Copenhagen, DK-2100, Denmark}
\affil[*]{luku@dtu.dk}
\affil[*]{markus.strobl@psi.ch}

\begin{document}

\flushbottom
\maketitle

\thispagestyle{empty}

%\noindent Please note: Abbreviations should be introduced at the first mention in the main text – no abbreviations lists. Suggested structure of main text (not enforced) is provided below.

\begin{abstract}
Obtaining a complete picture of local processes still poses a significant challenge in battery research. Here we demonstrate an \textit{in-situ} combination of multimodal neutron imaging with neutron diffraction for spatially resolved \textit{operando} observations of the lithiation-delithiation of a graphite electrode in a Li-ion battery cell. Throughout the lithiation-delithiation process we image the Li distribution based on the local beam attenuation. Simultaneously, we observe the development of the lithiated graphite phases as a function of cycling time and electrode thickness and integral throughout its volume by diffraction contrast imaging and diffraction, respectively.
While the conventional imaging data allows to observe the Li uptake in graphite already during the formation of the solid electrolyte interphase, diffraction indicates the onset and development of the Li insertion/extraction globally, which supports the local structural transformation observations by diffraction contrast imaging.
\end{abstract}

\section*{Introduction}

Rechargeable lithium-ion batteries are the most popular type of energy storage device with an increasing demand on the market, either for portable electronic devices, electric vehicles or grid energy storage. 
Rechargeable lithium-ion batteries are complex electrochemical devices governed by redox reactions, which in turn affect the crystallographic structure and the micro-structure of the utilised electrodes. 
Understanding phase evolution and the link with the \ce{Li} distribution in active electrode materials can provide insights into the electrochemical and physical processes and properties of batteries with the aim to support designing improved electrode and electrolyte materials.
A major challenge in investigating batteries is accessing the wide range of relevant scales, i.e. from electron and ion transport level to electrochemical cell scale.
Another significant challenge arises from the dynamism of the system, which is required to be enclosed, often leading to the necessity of designing special cells for \textit{in-situ} and \textit{operando} investigations.
In order to address the multi-scale spatial and temporal components, here we report an \textit{operando} investigation of a custom-made lithium-ion cell applying simultaneous multimodal neutron imaging and diffraction. 

While neutron and X-ray diffraction are well suited to observe lithiation and delithiation based on crystallographic observations~\cite{Dolotko2014, Vadlamani2014, TRUCANO1975, Senyshyn2012, Johnsen2013,Zhao2020}, neutrons, in contrast to X-rays, provide high visibility for Li, which allows direct observations of \ce{Li+} transport~\cite{Waldmann2016, Kardjilov2011, Siegel2011}.
Diffraction imaging thereby enables additional spatial resolution~\cite{Senyshyn2015}. However, to date, only few \textit{in-situ}/\textit{operando} studies employ a combination of such characterisation techniques to understand the fundamental processes in Li-ion batteries~\cite{Zhao2020, Senyshyn2012, Zhou2016, Chih-Jung2016, Saravanan2017, Wen2018, Ziesche2020}. Zhou~et~al., in a recent study~\cite{Zhou2016}, demonstrated the potential of conventional attenuation contrast neutron imaging in conjunction with neutron diffraction measurements by investigating the \ce{Li/Li+} transport and heterogeneity in a Li-ion battery on multiple length scales. However, neutron imaging and diffraction had to be performed consecutively, and thus, separately.
Furthermore, Kino~et~al.~\cite{Kino2015} employed diffraction contrast neutron imaging~\cite{WORACEK2018}, where diffraction provides a signal in wavelength-dependent transmission imaging, to study the local crystallographic phase transformations in the electrodes of a commercial battery. However, the envisaged advantage of direct spatial resolution in conjunction with the contained diffraction information remained limited due to neutron flux and resolution limitations. 
Here, in contrast, we do not only demonstrate an \textit{in-situ} combination of attenuation contrast imaging and diffraction, but we additionally exploit simultaneous diffraction contrast imaging to investigate a model battery system in \textit{operando}. 
The attenuation contrast of neutron imaging provides local measures of the lithium uptake of and release from the graphite electrode during the electrochemical cycle. Hence, it does, in principle, allow observations of various states of lithiation, particularly of the lithium transport across the electrode. 
However, neutron attenuation is also very sensitive to the electrolyte and is regularly used for imaging the variations of the electrolyte fill level in batteries~\cite{Lanz2001,Habedank2019, WEYDANZ2018, Zhao2020}. Thus, this implies potential bias in the observation of lithium concentrations, in particular, if the electrolyte is not deuterated~\cite{Zhou2016, Vadlamani2014}, as is illustrated by examples from pre-studies in the Supporting Information~\cite{Shinohara2016} (Figures~S3 and S4).
Neutron diffraction allows the characterisation of the lithiation states and phase transitions of the graphite electrode during the cycling process~\cite{Zhao2020, Dolotko2014, Vadlamani2014, TRUCANO1975} and therefore complements the spatially resolved information obtained by imaging. Due to our \textit{in-situ} approach, the respective information can be directly correlated. However, the diffraction information is not spatially resolved. Therefore, here we utilise an additional imaging modality, enabled by performing wavelength resolved imaging providing diffraction contrast through Bragg edges in the transmission spectra. 
The results of diffraction contrast neutron imaging provide spatio-temporal maps of the crystallographic phase evolution. Again, these results can be correlated and the findings of diffraction contrast imaging are supported by the integral \textit{in-situ} diffraction data.
The combined results of these simultaneously applied characterisation techniques offer a comprehensive view of the processes in the electrode on multiple scales, from \AA ngstr\"{o}ms to millimetres.
Thus, this study is suited to establish a proof-of-principle with wide implications for studying complex processes including those in energy storage and conversion devices and batteries in particular. It is furthermore of utmost relevance for the design of instruments for sophisticated multimodal studies at advanced neutron sources such as the European Spallation Source and in particular the Spallation Neutron Source in the United States and its planned second target station.

\section*{Methods}

The working principles and key capabilities of conventional neutron imaging and diffraction applied to Li-ion batteries have been reported earlier~\cite{Lanz2001,Riley2010, Siegel2011, Butler2011, Owejan2012, Same2012,Senyshyn2012,Siegel2013, Kino2015, Michalak2015, Zhou2016, Zhang2017,Song2019, Nie2019, Ziesche2020}.
Relevant details regarding the multiple techniques applied here, and an illustration of the integrative experimental set-up utilised (Figure~S1), are presented in the Supporting Information~\cite{Boin2012a}.
The experiments reported here were performed at the SENJU instrument~\cite{Ohhara2016,Kawasaki2014} on the beamline BL18 of the Materials and Life Science Experimental Facility of the Japan Proton Accelerator Research Complex (J-PARC). For this experiment, the instrument, a single crystal diffractometer by design, was retrofitted for multimodal investigations with an additional time-of-flight imaging detector developed at UC Berkeley~\cite{TREMSIN2020} to enable wavelength-resolved high-resolution imaging at pulsed neutron beams~\cite{WORACEK2018}.

\begin{figure}
    \centering
    \includegraphics[scale=0.46, keepaspectratio]{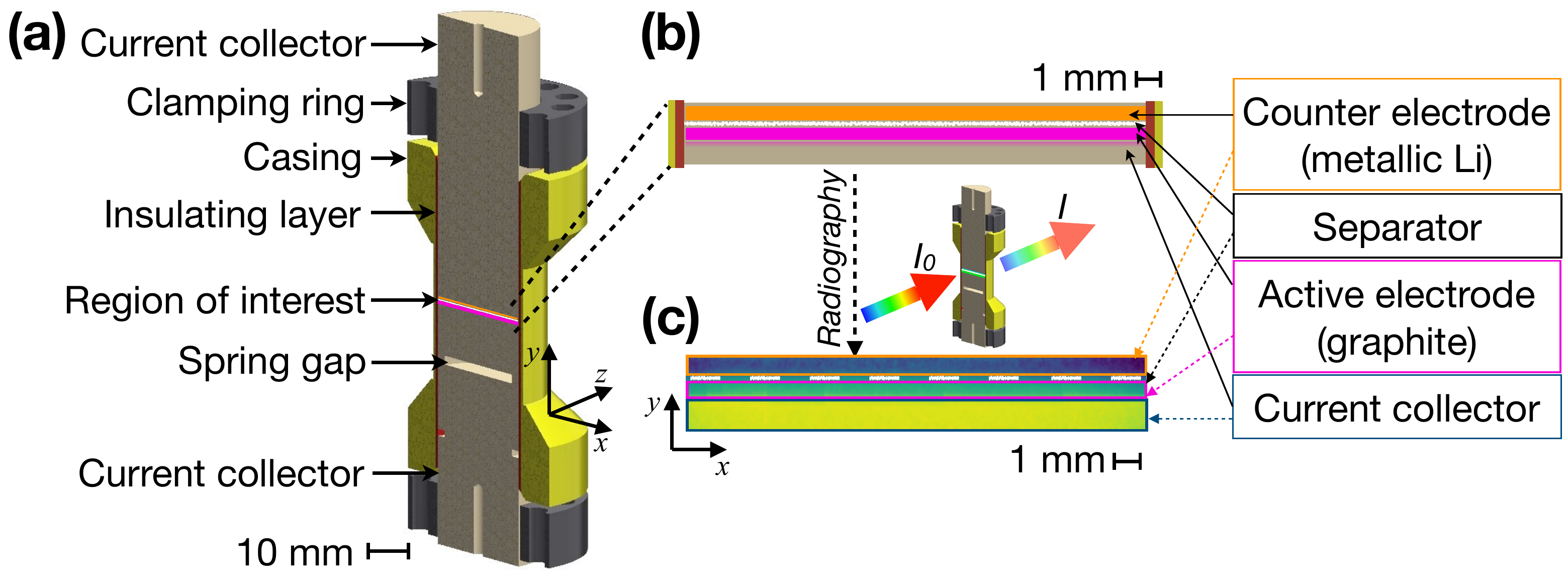}
    \caption{(a)~Illustration of the custom-made model battery cell optimized for neutron imaging investigations. (b)~A zoom in the cell illustration of the region exposed to neutrons. (c)~Neutron attenuation coefficient image of the region exposed to neutrons. The cell components and the components exposed to neutrons are indicated in the image.}
    \label{fig:cell}
\end{figure}

Figure~\ref{fig:cell}a depicts the electrochemical cell configuration designed and optimised for the in-plane neutron imaging and diffraction investigations, and the key components are indicated. 
The region of interest in the Li/graphite cell containing the active components in a stratified configuration is enlarged in Figure~\ref{fig:cell}b and juxtaposed to a corresponding attenuation contrast neutron image of the corresponding in Figure~\ref{fig:cell}c. 
The active electrode is a graphite-based pellet with 400~$\mu$m thickness (projected height), 16~mm diameter with an active mass of 0.1297~g. The separator is a glass microfiber filter with 260~$\mu$m (uncompressed) thickness. The counter electrode consists of a metallic Li disk.
The regions of the corresponding layers are indicated in the  representative colour coded neutron image in Figure~\ref{fig:cell}c for straightforward identification.

\section*{Results}

\begin{figure}
    \centering
    \includegraphics[scale=0.25, keepaspectratio]{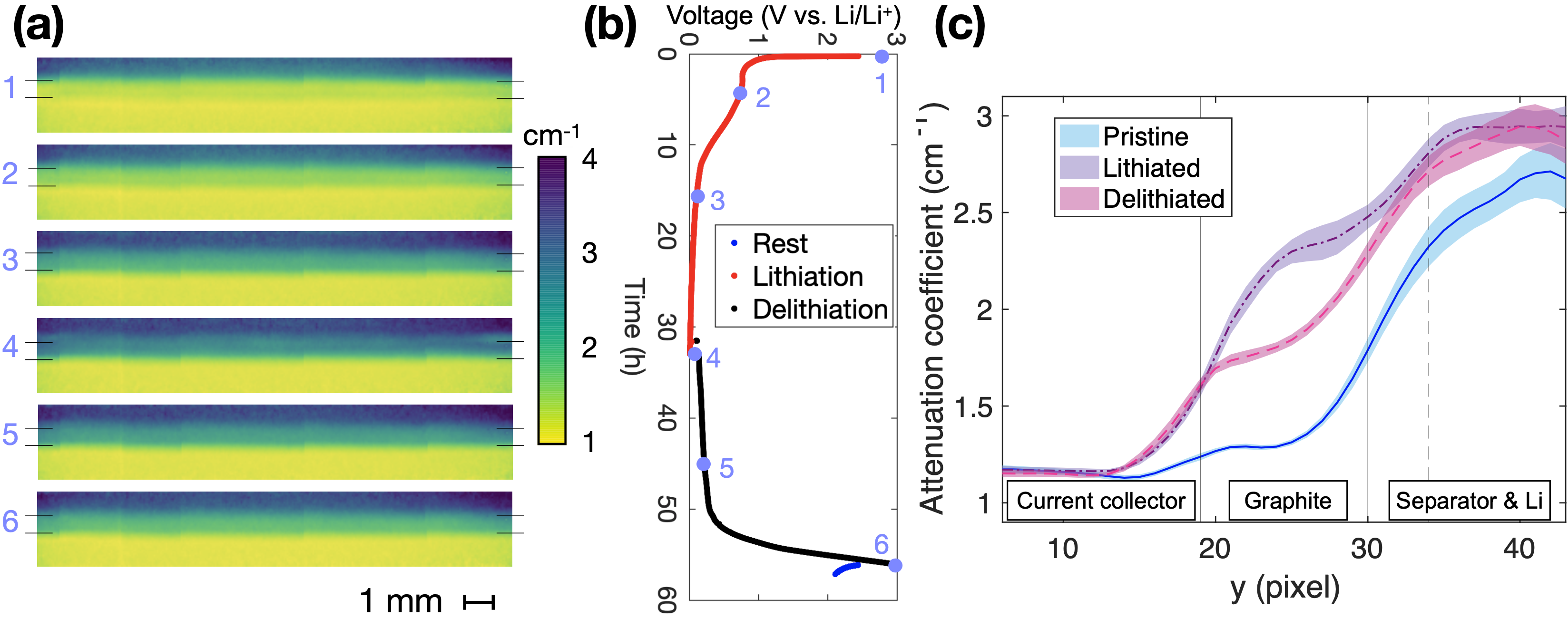}
    \caption{(a)~Attenuation coefficient contrast images at different stages during the electrochemical cycle. On the margins the lines indicate the graphite electrode position. (b)~Voltage profile versus time for the first cycle. The corresponding stages of the attenuation contrast images are indicated with circles and numbered. (c)~Average of \textit{y}-profiles and their standard deviation across the \textit{x}-direction for the pristine, lithiated and delithiated states. The vertical lines roughly demarcate the components. }
    \label{fig:initialres}
\end{figure}

Six representative normalised attenuation contrast images at various stages of the first lithiation-delithiation cycle are depicted in Figure~\ref{fig:initialres}a. Figure~\ref{fig:initialres}b displays the voltage profile of the electrochemical process, indicating the conditions at which the images in (a) are taken.
The recorded and reduced attenuation contrast images are normalised by the path length through the cylindrical sample. 
The images illustrate the detected increase in attenuation during lithiation which is correlated with the \ce{Li+} migration and insertion into the graphite electrode. Accordingly, a decrease of attenuation is detected during delithiation due to the reversal of the process. 
In Figure~\ref{fig:initialres}c, representative vertical line profiles of the attenuation contrast images are plotted based on the average behaviour across the width of the electrode for the pristine, most lithiated and delithiated states of the battery. The shaded area accompanying these line profiles illustrate the deviations of individual line profiles across the electrode from the average. This indicates that the Li distribution is rather homogeneous across the width of the electrode throughout the process. The line profiles are further suited to identify the interfaces of the different layers of the stratified structure. 
The effect of Li content build up in the graphite, and its different concentration from high values at the separator to much lower values towards the current collector in the most lithiated state are clearly distinguished. 
In order to enable a straightforward assessment of the lithiation and delithiation of the graphite electrode throughout the cycle, we utilise the fact that the behaviour is homogeneous across the electrode width, which allows horizontal (x-axis) integration.
Thus, every time bin of the observed process can be represented by a profile like shown in Figure~\ref{fig:initialres}c, which enables us to produce maps of the process in which one dimension is the vertical image direction, and the other is time.

\begin{figure*}[tb!]
    \centering
    \includegraphics[scale=0.38, keepaspectratio]{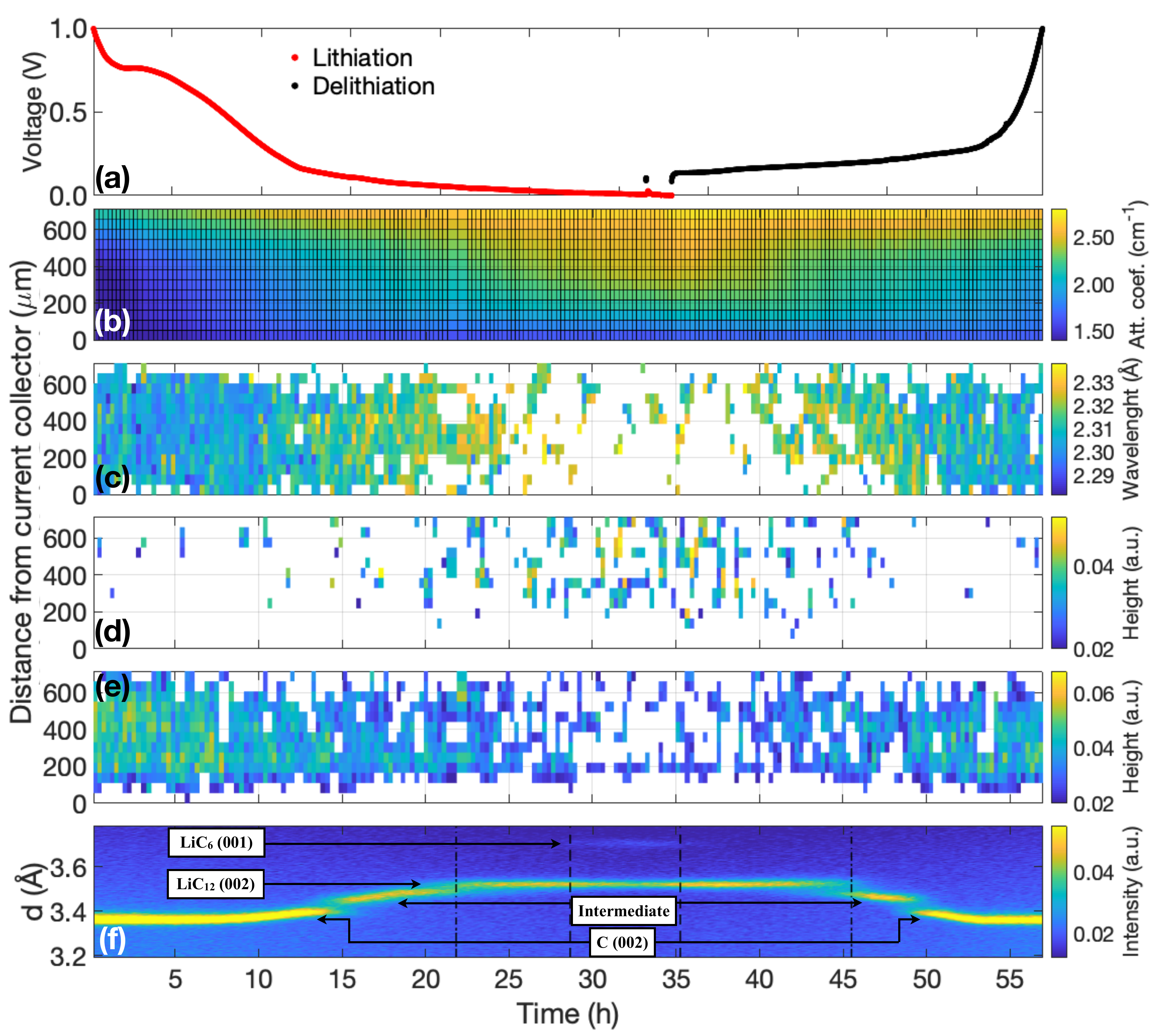}
    \caption{(a)~Voltage profile versus time for the first lithiation-delithiation cycle. (b)~Attenuation coefficient map of graphite electrode. (c)~Map of Bragg edge position for two adjacent edges of C and \ce{LiC12}. (d)~Map of a \ce{LiC6} Bragg edge height. (e)~Map of Bragg edge height for two overlapping edges of C and \ce{LiC12}.
    (b-e)~Each point is obtained by averaging the data over the pellet along the \textit{x}-direction and plotted according to the cycling time. The distance on \textit{y}-axis is given as: 0 is closer to the current collector and as it increases it gets closer to the separator. (f)~Neutron diffraction pattern during the cycling time.}
    \label{fig:results}
\end{figure*}

Figure~\ref{fig:results} presents such maps of the attenuation (Figure~\ref{fig:results}b) and three different diffraction contrast modalities extracted from the wavelength-dependent images (Figure~\ref{fig:results}c-e) alongside with the electrochemical process profile of the galvanostatic cycling (Figure~\ref{fig:results}a) and the results of the \textit{in-situ} diffraction measurement (Figure~\ref{fig:results}f), where the vertical axis is crystallographic lattice spacing. 
The diffraction contrast images are based on the analysis of different Bragg edges in the attenuation spectrum~\cite{Carminati2020}. 
The analyses focused on three narrow regions in the spectrum. Namely in Figure~\ref{fig:results}c the local Bragg edge position is retrieved for an edge at approximately 2.32~\AA. The region contains two edges, where a position at nominally $\lambda=2.31$~\AA~is indicative of graphite, i.e. the (1~1~2) reflection of graphite 2H, \textit{P}6$_3$/\textit{mmc} space group~\cite{TRUCANO1975}, and a weaker edge at nominally $\lambda=2.33$~\AA~is related to the (3~0~2) reflection of the \ce{LiC12} \textit{P}6/\textit{mmm} space group~\cite{Vadlamani2014}.
In addition, the height of an edge found at $\lambda=2.36$~\AA~indicates the lithiated phase \ce{LiC6} by the (3~0~1) reflection of \textit{P}6/\textit{mmm} space group~\cite{Dolotko2014} in Figure~\ref{fig:results}d. Finally, in Figure~\ref{fig:results}e the height of an edge at $\lambda=2.46$~\AA~where the (1~1~0) reflection of graphite~\cite{TRUCANO1975} and the weaker (3~0~0) reflection of \ce{LiC12}~\cite{Vadlamani2014} overlap is a measure of the presence of these two phases. While detection of this edge hints at the presence of at least one of these two phases, the graphite edge is stronger, thus a loss in edge height indicates the formation of \ce{LiC12}, and the disappearance of the edge indicates the formation of \ce{LiC6}. Therefore, the analyses of these edges should provide full local information of the phases present in the electrode.

\begin{figure*}[tb!]
    \centering
    \includegraphics[scale=0.41, keepaspectratio]{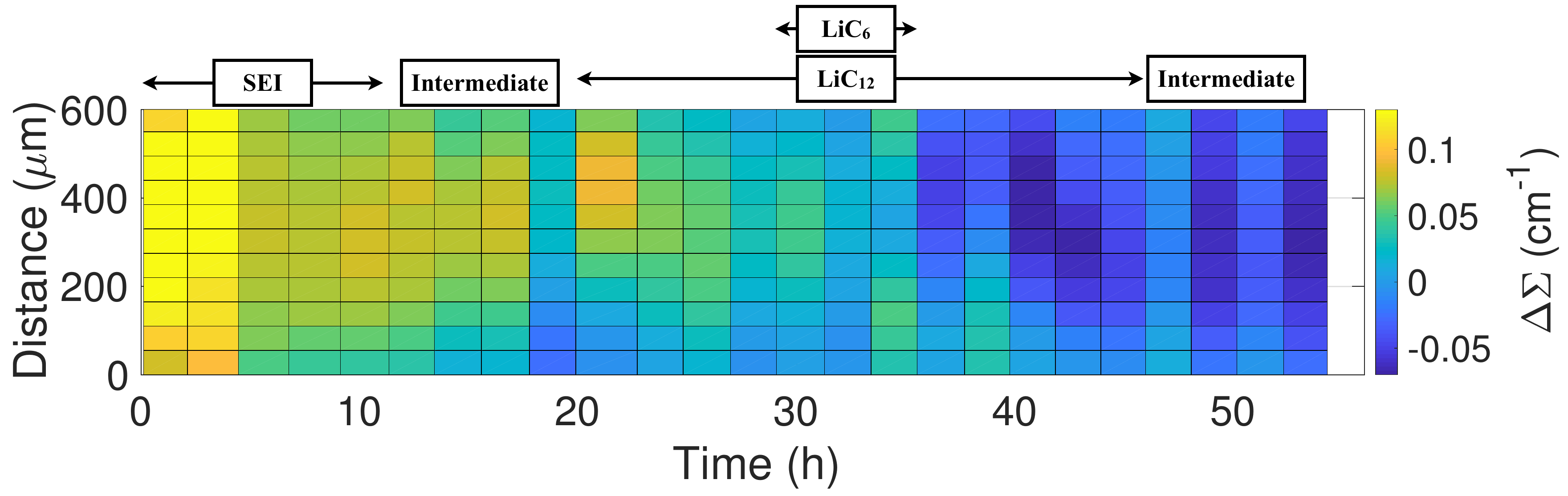}
    \caption{Map of relative attenuation change with increment of two hours; increasing attenuation with regards to the previous 2 hours is displayed in brighter colors of green to yellow, decreasing relative attenuation in darker, blue color range.}
    \label{fig:difres}
\end{figure*}

In Figure~\ref{fig:results}a the voltage profile from Figure~\ref{fig:initialres}b is replotted on the same time scale as the other results to enable correlation of the measured data with the process status. 
During approximately the first 10 hours of lithiation, the shoulder appearing at voltages between 0.8-0.2~V~vs.~\ce{Li/Li+} can be related to the formation of the solid electrolyte interphase (SEI)~\cite{An2016,VERMA2010}, a thin passivation layer formed on the graphite surface during the first cycle due to electrochemical decomposition of electrolyte. 
This is clearly reflected in Figure~\ref{fig:results}b by a smooth increase of attenuation which is almost uniform throughout the graphite electrode.
It is accompanied by a commencing significant increase of attenuation in the separator and Li electrode region, which appears to hint to an increasing amount of electrolyte and is hypothesised to be due to electrolyte replacing the consumed Li metal in this region.
The diffraction contrast images (Figure~\ref{fig:results}c-e), and the global diffraction results (Figure~\ref{fig:results}f) display no changes in this time span, underlining that no \ce{Li+} is inserted in the graphite during this process period up to approximately 10~hours into the process.
Figure~\ref{fig:difres}, which displays the changes in attenuation with an increment of about two hours presents the subtle changes in the lithiation process, indicating that the SEI formation indeed possesses a slight gradient in time from the separator interface towards deeper in the electrode and slows down towards a minimum at approximately 10 hours.

Beyond that time, \ce{Li+} ingress into the graphite is detected by diffraction (Figure~\ref{fig:results}f) and diffraction contrast imaging where the graphite Bragg edge at 2.31~\AA~starts shifting towards the \ce{LiC12} edge at 2.33~\AA~(Figure~\ref{fig:results}c) and thus the crystallographic phase evolution determined by the lithium insertion begins at a potential of 0.2~V~vs.~\ce{Li/Li+}~\cite{An2016}. 
A short uniform acceleration in attenuation increase, i.e. Li uptake, is measured at this point in time, returning again to very small values around 15 hours and shifting towards the separator interface (Figure~\ref{fig:difres}). The \ce{Li+}-insertion strongly accelerates with the formation of \ce{LiC12} after approximately 20 hours (Figure~\ref{fig:difres}), which is clearly implied not only by diffraction (Figure~\ref{fig:results}f), but also by diffraction contrast imaging in Figure~\ref{fig:results}c and e. 
Further, it slows down while moving deeper into the active electrode. Meanwhile, a considerable gradient in Li-content from the separator towards the current collector~\cite{Yao2019} has formed, evident in the attenuation contrast plot. After approximately 28 hours, the \ce{LiC6} content reaches significant values (Figure~\ref{fig:results}d, f), but the growth in Li concentration is rather moderate at that time and formation appears to start relatively centred in the layer (Figure~\ref{fig:difres}).
Already at about 25 hours, before \ce{LiC6} formation commences, the \ce{LiC12} signal in the diffraction contrast panel Figure~\ref{fig:results}c vanishes, which might partially be due to the large attenuation of the signal. Beyond 28 hours, also the diffraction contrast in Figure~\ref{fig:results}e indicates a significant loss of \ce{LiC12} phase, in favour of \ce{LiC6}, while the total Li-content hardly increases further (Figure~\ref{fig:difres}) as the graphite lithiation process was not complete. Despite the high attenuation, the diffraction contrast signal of \ce{LiC6} is partially detected and displayed in panel (d) of Figure~\ref{fig:results}. 
Upon delithiation from 33~hours, the process is reverted and the Li-gradient is decreasing due to \ce{Li+}-extraction foremost at the interface. Interestingly the electrodes appear to move downwards (against the spring) by approximately one pixel (55~$\mu$m) during the first hours of delithiation, indicated by apparently increased attenuation at the peripheries of the graphite electrode (Figure~\ref{fig:difres}). 
Between 36 and 46 hours, the peak of delithiation moves from the separator interface deeper into the pellet, while it reaches a maximum between 40 and 42~hours (Figure~\ref{fig:difres}), the onset of ultimately decreasing \ce{LiC12} phase (Figure~\ref{fig:results}). Delithiation beyond 45~hours is slower and relatively homogeneous throughout the thickness after the lithiation gradient has vanished. The fully delithiated state at 55~hours resembles the state of full SEI formation at about 12~hours.
Overall, the SEI formation contributes more in total to the attenuation increase in the graphite electrode than the \ce{Li+}-insertion, which is on par with the SEI only at locations of strongest insertion. These results seem to indicate that during the delithiation all inserted \ce{Li+} are extracted again, and only the SEI remains in the active electrode (compare Figure~\ref{fig:initialres}c).

\section*{Discussion}

In summary, we have not only demonstrated that multimodal neutron imaging, namely attenuation and diffraction contrast imaging, can be combined with \textit{in-situ} diffraction in a single \textit{operando} experiment, but the complementary information obtained provides unique insights for battery research. 
Apart from macroscopic evaluations like the electrode volumetric expansion, the redistribution of electrolyte, gas bubbles, and positional rearrangement (Figures S3-S5 in the Supporting Information), the attenuation contrast enables observations of the SEI formation. 
This information can be clearly offset from attenuation due to \ce{Li+}-insertion based on the diffraction information, both in diffraction contrast imaging and diffraction. 
The imaging methods further enable the localisation of these processes, while diffraction can act as an integral control (compare to Figure S6 in the Supporting Information). 
This way, a uniform formation of the SEI could be observed throughout the graphite electrode and contributes the largest fraction of the attenuation increase in this regime. It has been directly observed that, in contrast, the \ce{Li+}-insertion creates a gradient in Li-content, which is higher towards the separator and \ce{Li+}-insertion rates appear highest during \ce{LiC12} formation in our active electrode. 
Further, it could be observed that basically all inserted Li was extracted during the delithiation process and the state at the end of SEI formation could be reproduced. 
The results suggest that the SEI formation was largely complete and did not advance significantly when Li was inserted. 
The diffraction imaging signal furthermore enables the exclusion of potential bias due to electrolyte movement and bubble formation as compared to sole attenuation contrast imaging and adds spatial resolution to phase transformation observations in contrast to diffraction. 
This enabled unique insights into the local SEI formation and \ce{Li+}-insertion dynamics, which are essential in understanding the lithiation process.

It is thus concluded, that this \textit{in-situ} process characterisation is superior to separate individual measurements with different techniques and holds great potential for battery studies and most likely also for several other fields of applied material research. Novel neutron sources like the European Spallation Source (ESS) and in particular the Spallation Neutron Source (SNS) with its second target station instrumentation currently under consideration are promising to enable sophisticated combinations of these techniques for advanced \textit{in-situ} and \textit{operando} studies.

\section*{Acknowledgements}

This project has received funding from DANSCATT in relation to travel expenses for the beamtime. This work was also supported by OP RDE, MEYS, under the project "European Spallation Source - participation of the Czech Republic - OP”, Reg. No. \linebreak CZ.02.1.01/0.0/0.0/16\_013/0001794. This experiment was performed at SENJU instrument at the BL18 beamline (Proposal No. 2018B0076) and pre-studies were performed at RADEN instrument at BL22 (Proposal No. 2017B0114) at J-PARC. Further, the authors thank Christian Baur and Sigita Trabesinger for helpful discussions on batteries and electrochemistry.

\section*{Author contributions statement}

M.-E.L. and R.E.J. planned the project with the help of L.T.K., M.S., R.W. and S.S.. M.-E.L. designed the cell and synthesised the active electrode with support from R.E.J.. The neutron data was acquired by M.-E.L. with the help of R.E.J., P.K.M.T., L.T.K., S.S., N.E., T.S., R.K. and A.S.T.. The image data was analysed by M.-E.L. and M.S. and the diffraction by P.K.M.T.. M.-E.L. and M.S. wrote the manuscript. The manuscript was revised with the help of all authors.

\section*{Additional information}

The data that support the findings of this study are available from the corresponding author upon reasonable request.

Supporting Information Available: 
Experimental methods; cell design and assembly; \textit{Operando} neutron investigation details; schematic illustration of the setup (Figure S1); diffraction contrast neutron imaging; data processing; and additional data and figures.

\newpage
\printbibliography

@article{Shinohara2016,
author = {Shinohara, T. and Kai, T. and Oikawa, K. and Segawa, M. and Harada, M. and Nakatani, T. and Ooi, M. and Aizawa, K. and Sato, H. and Kamiyama, T. and Yokota, H. and Sera, T. and Mochiki, K. and Kiyanagi, Y.},
doi = {https://doi.org/10.1088/1742-6596/746/1/012007},
issn = {1742-6588},
journal = {Journal of Physics: Conference Series},
number = {1},
pages = {12007},
title = {{Final design of the Energy-Resolved Neutron Imaging System 'RADEN' at J-PARC}},
url = {http://stacks.iop.org/1742-6596/746/i=1/a=012007?key=crossref.275cb7179d0846bf1ccd928d1d04e19e},
volume = {746},
year = {2016}
}

@article{TREMSIN2020,
title = {{Unique capabilities and applications of Microchannel Plate (MCP) detectors with Medipix/Timepix readout}},
journal = {Radiation Measurements},
volume = {130},
pages = {106228},
year = {2020},
issn = {1350-4487},
doi = {https://doi.org/10.1016/j.radmeas.2019.106228},
url = {http://www.sciencedirect.com/science/article/pii/S1350448719305141},
author = {A.S. Tremsin and J.V. Vallerga}
}

@article{TRUCANO1975,
author = {Trucano, P. and Chen, R.},
journal = {Nature},
month = {11},
pages = {136},
publisher = {Nature Publishing Group},
title = {{Structure of graphite by neutron diffraction}},
url = {http://dx.doi.org/10.1038/258136a0 http://10.0.4.14/258136a0},
doi = {https://doi.org/10.1038/258136a0},
volume = {258},
year = {1975}
}

@article{Vadlamani2014,
author = {Vadlamani, B. and An, K. and Jagannathan, M. and Chandran, K. S. Ravi},
doi = {https://doi.org/10.1149/2.0951410jes},
issn = {0013-4651},
journal = {Journal of The Electrochemical Society},
month = {07},
number = {10},
pages = {A1731--A1741},
publisher = {The Electrochemical Society},
title = {{An In-Situ Electrochemical Cell for Neutron Diffraction Studies of Phase Transitions in Small Volume Electrodes of Li-Ion Batteries}},
url = {http://jes.ecsdl.org/lookup/doi/10.1149/2.0951410jes},
volume = {161},
year = {2014}
}

@article{Boin2012a,
author = {Boin, M.},
journal = {J. Appl. Crystallogr.},
title = {Nxs: A program library for neutron cross section calculations},
doi = {https://doi.org/10.1107/S0021889812016056},
volume = {45}, 
pages = {603-607},
year = {2012}
}

@article{Kino2015,
author = {Kino, Koichi and Yonemura, Masao and Kiyanagi, Yoshiaki and Ishikawa, Yoshihisa and Parker, Joseph Don and Tanimori, Toru and Kamiyama, Takashi},
doi = {https://doi.org/10.1016/j.phpro.2015.07.087},
journal = {Physics Procedia},
number = {69},
pages = {612--618},
title = {{First Imaging Experiment of a Lithium Ion Battery by a Pulsed Neutron Beam at J-PARC/MLF/BL09}},
url = {https://www.sciencedirect.com/science/article/pii/S1875389215006975},
volume = {69},
year = {2015}
}

@article{Ohhara2016,
author = {Ohhara, Takashi and Kiyanagi, Ryoji and Oikawa, Kenichi and Kaneko, Koji and Kawasaki, Takuro and Tamura, Itaru and Nakao, Akiko and Hanashima, Takayasu and Munakata, Koji and Moyoshi, Taketo and Kuroda, Tetsuya and Kimura, Hiroyuki and Sakakura, Terutoshi and Lee, Chang-Hee and Takahashi, Miwako and Ohshima, Ken-ichi and Kiyotani, Tamiko and Noda, Yukio and Arai, Masatoshi},
title = {{SENJU: a new time-of-flight single-crystal neutron diffractometer at J-PARC}},
journal = {Journal of Applied Crystallography},
year = {2016},
volume = {49},
number = {1},
pages = {120--127},
month = {02},
doi = {https://doi.org/10.1107/S1600576715022943},
url = {https://doi.org/10.1107/S1600576715022943}
}

@article{Dolotko2014,
title = {{Understanding structural changes in NMC Li-ion cells by in situ neutron diffraction}},
journal = {Journal of Power Sources},
volume = {255},
pages = {197--203},
year = {2014},
issn = {0378-7753},
doi = {https://doi.org/10.1016/j.jpowsour.2014.01.010},
url = {http://www.sciencedirect.com/science/article/pii/S0378775314000214},
author = {O. Dolotko and A. Senyshyn and M.J. Muehlbauer and K. Nikolowski and H. Ehrenberg}
}

@article{Zhou2016,
author = {Zhou, Hui and An, Ke and Allu, Srikanth and Pannala, Sreekanth and Li, Jianlin and Bilheux, Hassina Z. and Martha, Surendra K. and Nanda, Jagjit},
title = {Probing Multiscale Transport and Inhomogeneity in a Lithium-Ion Pouch Cell Using In Situ Neutron Methods},
journal = {ACS Energy Letters},
volume = {1},
number = {5},
pages = {981-986},
year = {2016},
doi = {https://doi.org/10.1021/acsenergylett.6b00353},
URL = {https://doi.org/10.1021/acsenergylett.6b00353}
}

@article{Saravanan2017,
	Author = {Kuppan, Saravanan and Xu, Yahong and Liu, Yijin and Chen, Guoying},
	Journal = {Nature Communications},
	Number = {1},
	Pages = {14309},
	Title = {Phase transformation mechanism in lithium manganese nickel oxide revealed by single-crystal hard X-ray microscopy},
	doi = {https://doi.org/10.1038/ncomms14309},
	Volume = {8},
	Year = {2017}}

@article{Chih-Jung2016,
author = {Chen, Chih-Jung and Pang, Wei Kong and Mori, Tatsuhiro and Peterson, Vanessa K. and Sharma, Neeraj and Lee, Po-Han and Wu, She-huang and Wang, Chun-Chieh and Song, Yen-Fang and Liu, Ru-Shi},
title = {The Origin of Capacity Fade in the \ce{Li2MnO3}$\cdot$\ce{LiMO2} (M = Li, Ni, Co, Mn) Microsphere Positive Electrode: An Operando Neutron Diffraction and Transmission X-ray Microscopy Study},
journal = {Journal of the American Chemical Society},
volume = {138},
number = {28},
pages = {8824-8833},
year = {2016},
doi = {https://doi.org/10.1021/jacs.6b03932},
URL = { 
        https://doi.org/10.1021/jacs.6b03932
}
}

@ARTICLE{Wen2018,
AUTHOR={Zhu, Wen and Liu, Dongqiang and Paolella, Andrea and Gagnon, Catherine and Gari\'{e}py, Vincent and Vijh, Ashok and Zaghib, Karim},   
TITLE={Application of Operando X-ray Diffraction and Raman Spectroscopies in Elucidating the Behavior of Cathode in Lithium-Ion Batteries},      
JOURNAL={Frontiers in Energy Research},      
VOLUME={6},      
PAGES={66},     
YEAR={2018},      
URL={https://www.frontiersin.org/article/10.3389/fenrg.2018.00066},       
DOI={https://doi.org/10.3389/fenrg.2018.00066},      
ISSN={2296-598X}
}

@article{Ziesche2020,
	Author = {Ziesche, Ralf F. and Arlt, Tobias and Finegan, Donal P. and Heenan, Thomas M. M. and Tengattini, Alessandro and Baum, Daniel and Kardjilov, Nikolay and Mark{\"o}tter, Henning and Manke, Ingo and Kockelmann, Winfried and Brett, Dan J. L. and Shearing, Paul R.},
	Journal = {Nature Communications},
	Number = {1},
	Pages = {777},
	Title = {4D imaging of lithium-batteries using correlative neutron and X-ray tomography with a virtual unrolling technique},
	doi = {https://doi.org/10.1038/s41467-019-13943-3},
	Volume = {11},
	Year = {2020}}

@article{Siegel2011,
author = {Siegel, Jason B and Lin, Xinfan and Stefanopoulou, Anna G and Hussey, Daniel S and Jacobson, David L and Gorsich, David},
doi = {https://doi.org/10.1149/1.3566341},
isbn = {0013-4651},
issn = {00134651},
journal = {Journal of The Electrochemical Society},
number = {5},
pages = {A523},
title = {{Neutron Imaging of Lithium Concentration in LFP Pouch Cell Battery}},
volume = {158},
year = {2011}
}

@article{Same2012,
author = {Same, Adam and Battaglia, Vincent and Tang, Hong-Yue and Park, Jae Wan},
doi = {https://doi.org/10.1007/s10800-011-0363-3},
issn = {0021-891X},
journal = {Journal of Applied Electrochemistry},
month = {jan},
number = {1},
pages = {1--9},
publisher = {Springer Netherlands},
title = {{In situ neutron radiography analysis of graphite/NCA lithium-ion battery during overcharge}},
url = {http://link.springer.com/10.1007/s10800-011-0363-3},
volume = {42},
year = {2012}
}

@article{Michalak2015,
author = {Michalak, Barbara and Sommer, Heino and Mannes, David and Kaestner, Anders and Brezesinski, Torsten and Janek, J{\"{u}}rgen},
journal = {Scientific Reports},
month = {oct},
pages = {15627},
publisher = {The Author(s)},
title = {{Gas Evolution in Operating Lithium-Ion Batteries Studied In Situ by Neutron Imaging}},
doi = {https://doi.org/10.1038/srep15627},
url = {http://dx.doi.org/10.1038/srep15627 http://10.0.4.14/srep15627 https://www.nature.com/articles/srep15627{\#}supplementary-information},
volume = {5},
year = {2015}
}

@article{Lanz2001,
author = {Lanz, Martin and Lehmann, Eberhard and Imhof, Roman and Exnar, Ivan and Nov{\'{a}}k, Petr},
doi = {https://doi.org/10.1016/S0378-7753(01)00706-6},
issn = {03787753},
journal = {Journal of Power Sources},
month = {oct},
number = {2},
pages = {177--181},
title = {{In situ neutron radiography of lithium-ion batteries during charge/discharge cycling}},
url = {https://linkinghub.elsevier.com/retrieve/pii/S0378775301007066},
volume = {101},
year = {2001}
}

@article{Siegel2013,
author = {Siegel, Jason B. and Stefanopoulou, Anna G. and Hagans, Patrick and Ding, Yi and Gorsich, David},
doi = {https://doi.org/10.1149/2.011308jes},
issn = {0013-4651},
journal = {Journal of The Electrochemical Society},
month = {apr},
number = {8},
pages = {A1031--A1038},
publisher = {The Electrochemical Society},
title = {{Expansion of Lithium Ion Pouch Cell Batteries: Observations from Neutron Imaging}},
url = {http://jes.ecsdl.org/lookup/doi/10.1149/2.011308jes},
volume = {160},
year = {2013}
}

@inproceedings{Riley2010,
author = {Riley, Grant V. and Hussey, Daniel S. and Jacobson, David},
booktitle = {ECS Transactions},
doi = {https://doi.org/10.1149/1.3414005},
issn = {1938-5862},
month = {apr},
number = {35},
pages = {75--83},
publisher = {The Electrochemical Society},
title = {{In Situ Neutron Imaging Of Alkaline and Lithium Batteries}},
url = {http://ecst.ecsdl.org/cgi/doi/10.1149/1.3414005},
year = {2010}
}

@article{Owejan2012,
author = {Owejan, Jon P. and Gagliardo, Jeffrey J. and Harris, Stephen J. and Wang, Howard and Hussey, Daniel S. and Jacobson, David L.},
doi = {https://doi.org/10.1016/J.ELECTACTA.2012.01.047},
issn = {0013-4686},
journal = {Electrochimica Acta},
month = {apr},
pages = {94--99},
publisher = {Pergamon},
title = {{Direct measurement of lithium transport in graphite electrodes using neutrons}},
url = {https://www.sciencedirect.com/science/article/pii/S001346861200076X?via{\%}3Dihub},
volume = {66},
year = {2012}
}

@article{Zhang2017,
author = {Zhang, Y. and Chandran, K. S. Ravi and Jagannathan, M. and Bilheux, H. Z. and Bilheux, J. C.},
doi = {https://doi.org/10.1149/2.0051702jes},
issn = {0013-4651},
journal = {Journal of The Electrochemical Society},
month = {jan},
number = {2},
pages = {A28--A38},
publisher = {The Electrochemical Society},
title = {{The Nature of Electrochemical Delithiation of Li-Mg Alloy Electrodes: Neutron Computed Tomography and Analytical Modeling of Li Diffusion and Delithiation Phenomenon}},
url = {http://jes.ecsdl.org/lookup/doi/10.1149/2.0051702jes},
volume = {164},
year = {2017}
}

@article{Butler2011,
author = {Butler, Leslie G. and Schillinger, Burkhard and Ham, Kyungmin and Dobbins, Tabbetha A. and Liu, Ping and Vajo, John J.},
doi = {https://doi.org/10.1016/j.nima.2011.03.023},
issn = {01689002},
journal = {Nuclear Instruments and Methods in Physics Research Section A: Accelerators, Spectrometers, Detectors and Associated Equipment},
month = {sep},
number = {1},
pages = {320--328},
title = {{Neutron imaging of a commercial Li-ion battery during discharge: Application of monochromatic imaging and polychromatic dynamic tomography}},
url = {https://linkinghub.elsevier.com/retrieve/pii/S0168900211005638},
volume = {651},
year = {2011}
}

@article{Senyshyn2012,
author = {Senyshyn, A and M{\"{u}}hlbauer, M J and Nikolowski, K and Pirling, T and Ehrenberg, H},
doi = {https://doi.org/10.1016/j.jpowsour.2011.12.007},
isbn = {0378-7753},
issn = {03787753},
journal = {Journal of Power Sources},
pages = {126--129},
publisher = {Elsevier B.V.},
title = {{"In-operando" neutron scattering studies on Li-ion batteries}},
url = {http://dx.doi.org/10.1016/j.jpowsour.2011.12.007},
volume = {203},
year = {2012}
}

@article{Song2019,
	Author = {Song, Bohang and Dhiman, Indu and Carothers, John C. and Veith, Gabriel M. and Liu, Jue and Bilheux, Hassina Z. and Huq, Ashfia},
	Journal = {ACS Energy Letters},
	Month = {10},
	Number = {10},
	Pages = {2402--2408},
	Title = {Dynamic Lithium Distribution upon Dendrite Growth and Shorting Revealed by Operando Neutron Imaging},
	Volume = {4},
	Year = {2019},
	doi = {https://doi.org/10.1021/acsenergylett.9b01652}
	}

@article{Nie2019,
author = {Nie, Ziyang and McCormack, Patrick and Bilheux, Hassina Z. and Bilheux, Jean C. and Robinson, J. Pierce and Nanda, Jagjit and Koenig, Gary M.},
doi = {https://doi.org/10.1016/j.jpowsour.2019.02.075},
issn = {03787753},
journal = {Journal of Power Sources},
month = {apr},
pages = {127--136},
title = {{Probing lithiation and delithiation of thick sintered lithium-ion battery electrodes with neutron imaging}},
url = {https://linkinghub.elsevier.com/retrieve/pii/S0378775319302009},
volume = {419},
year = {2019}
}

@article{An2016,
title = {{The state of understanding of the lithium-ion-battery graphite solid electrolyte interphase (SEI) and its relationship to formation cycling}},
journal = {Carbon},
volume = {105},
pages = {52 - 76},
year = {2016},
issn = {0008-6223},
doi = {https://doi.org/10.1016/j.carbon.2016.04.008},
url = {http://www.sciencedirect.com/science/article/pii/S0008622316302676},
author = {Seong Jin An and Jianlin Li and Claus Daniel and Debasish Mohanty and Shrikant Nagpure and David L. Wood}
}

@article{Johnsen2013,
author = {Johnsen, Rune E. and Norby, Poul},
doi = {https://doi.org/10.1107/S0021889813022796},
issn = {0021-8898},
journal = {Journal of Applied Crystallography},
month = {dec},
number = {6},
pages = {1537--1543},
publisher = {International Union of Crystallography},
title = {{Capillary-based micro-battery cell for \textit{in-situ} X-ray powder diffraction studies of working batteries: a study of the initial intercalation and deintercalation of lithium into graphite}},
url = {http://scripts.iucr.org/cgi-bin/paper?S0021889813022796},
volume = {46},
year = {2013}
}

@article{Kawasaki2014,
title = {{Detector system of the SENJU single-crystal time-of-flight neutron diffractometer at J-PARC/MLF}},
journal = {Nuclear Instruments and Methods in Physics Research Section A: Accelerators, Spectrometers, Detectors and Associated Equipment},
volume = {735},
pages = {444 - 451},
year = {2014},
issn = {0168-9002},
doi = {https://doi.org/10.1016/j.nima.2013.09.057},
url = {http://www.sciencedirect.com/science/article/pii/S0168900213012916},
author = {Kawasaki, T. and Nakamura, T. and Toh, K. and Hosoya, T. and  Oikawa, K. and Ohhara,T. and Kiyanagi, R. and Ebine, M. and Birumachi, A. and Sakasai, K. and Soyama, K. and Katagiri, M.}
}

@article{WORACEK2018,
title = {{Diffraction in neutron imaging - A review}},
journal = {Nuclear Instruments and Methods in Physics Research Section A: Accelerators, Spectrometers, Detectors and Associated Equipment},
volume = {878},
pages = {141 - 158},
year = {2018},
note = {Radiation Imaging Techniques and Applications},
issn = {0168-9002},
doi = {https://doi.org/10.1016/j.nima.2017.07.040},
url = {http://www.sciencedirect.com/science/article/pii/S0168900217307817},
author = {Robin Woracek and Javier Santisteban and Anna Fedrigo and Markus Strobl}
}

@article{VERMA2010,
title = {{A review of the features and analyses of the solid electrolyte interphase in Li-ion batteries}},
journal = {Electrochimica Acta},
volume = {55},
number = {22},
pages = {6332 - 6341},
year = {2010},
issn = {0013-4686},
doi = {https://doi.org/10.1016/j.electacta.2010.05.072},
url = {http://www.sciencedirect.com/science/article/pii/S0013468610007747},
author = {Pallavi Verma and Pascal Maire and Petr Nov{\'{a}}k}
}

@article{Waldmann2016,
author = {Waldmann, Thomas and Iturrondobeitia, Amaia and Kasper, Michael and Ghanbari, Niloofar and Aguesse, Fr{\'{e}}d{\'{e}}ric and Bekaert, Emilie and Daniel, Lise and Genies, Sylvie and Gordon, Isabel Jim{\'{e}}nez and L{\"{o}}ble, Matthias W. and {De Vito}, Eric and Wohlfahrt-Mehrens, Margret},
doi = {https://doi.org/10.1149/2.1211609jes},
issn = {0013-4651},
journal = {Journal of The Electrochemical Society},
month = {aug},
number = {10},
pages = {A2149--A2164},
publisher = {The Electrochemical Society},
title = {{Review—Post-Mortem Analysis of Aged Lithium-Ion Batteries: Disassembly Methodology and Physico-Chemical Analysis Techniques}},
url = {http://jes.ecsdl.org/lookup/doi/10.1149/2.1211609jes},
volume = {163},
year = {2016}
}

@article{Kardjilov2011,
author = {Kardjilov, Nikolay and Manke, Ingo and Hilger, Andr{\'{e}} and Strobl, Markus and Banhart, John},
doi = {https://doi.org/10.1016/S1369-7021(11)70139-0},
issn = {13697021},
journal = {Materials Today},
month = {jun},
number = {6},
pages = {248--256},
title = {{Neutron imaging in materials science}},
url = {http://linkinghub.elsevier.com/retrieve/pii/S1369702111701390},
volume = {14},
year = {2011}
}

@article{Habedank2019,
	Author = {Habedank, Jan Bernd and G{\"u}nter, Florian J. and Billot, Nicolas and Gilles, Ralph and Neuwirth, Tobias and Reinhart, Gunther and Zaeh, Michael F.},
	Journal = {The International Journal of Advanced Manufacturing Technology},
	Number = {9},
	Pages = {2769--2778},
	Title = {Rapid electrolyte wetting of lithium-ion batteries containing laser structured electrodes: in situ visualization by neutron radiography},
	Volume = {102},
	doi = {https://doi.org/10.1007/s00170-019-03347-4},
	Year = {2019}}

@article{WEYDANZ2018,
title = {{Visualization of electrolyte filling process and influence of vacuum during filling for hard case prismatic lithium ion cells by neutron imaging to optimize the production process}},
journal = {Journal of Power Sources},
volume = {380},
pages = {126 - 134},
year = {2018},
issn = {0378-7753},
doi = {https://doi.org/10.1016/j.jpowsour.2018.01.081},
url = {http://www.sciencedirect.com/science/article/pii/S0378775318300818},
author = {W.J. Weydanz and H. Reisenweber and A. Gottschalk and M. Schulz and T. Knoche and G. Reinhart and M. Masuch and J. Franke and R. Gilles}
}

@article{Carminati2020,
author = {Carminati, Chiara and Strobl, Markus and Minniti, Triestino and Boillat, Pierre and Hovind, Jan and Morgano, Manuel and Holm Rod, Thomas and Polatidis, Efthymios and Valsecchi, Jacopo and Mannes, David and Kockelmann, Winfried and Kaestner, Anders},
title = {{Bragg-edge attenuation spectra at voxel level from 4D wavelength-resolved neutron tomography}},
journal = {Journal of Applied Crystallography},
year = {2020},
volume = {53},
number = {1},
pages = {188--196},
month = {Feb},
doi = {https://doi.org/10.1107/S1600576720000151},
url = {https://doi.org/10.1107/S1600576720000151}
}

@article{Zhao2020,
author = {Zhao, Enyue and Zhang, Zhi-Gang and Li, Xiyang and He, Lunhua and Yu, Xiqian and Li, Hong and Wang, Fangwei},
title = {{Neutron-based characterization techniques for lithium-ion battery research}},
journal = {Chinese Physics B},
year = {2020},
volume = {29},
number = {1},
pages = {018201},
doi = {https://doi.org/10.1107/S1600576720000151},
url = {DOI: 10.1088/1674-1056/ab5d07}
}

@article{Senyshyn2015,
author = {Senyshyn, A and M{\"{u}}hlbauer, M J and Dolotko, O and Hofmann, M and Ehrenberg, H},
doi = {https://doi.org/10.1038/srep18380},
issn = {20452322},
journal = {Scientific Reports},
number = {April},
pages = {1--9},
pmid = {26681110},
title = {{Homogeneity of lithium distribution in cylinder-type Li-ion batteries}},
url = {http://dx.doi.org/10.1038/srep18380},
year = {2015}
}

@article{Yao2019,
	Author = {Yao, Koffi P. C. and Okasinski, John S. and Kalaga, Kaushik and Shkrob, Ilya A. and Abraham, Daniel P.},
	Journal = {Energy Environ. Sci.},
	Pages = {656-665},
	Title = {Quantifying lithium concentration gradients in the graphite electrode of Li-ion cells using operando energy dispersive X-ray diffraction},
	Volume = {12},
	Year = {2019},
	doi = {https://doi.org/10.1039/C8EE02373E}
	}

\end{document}

\label{SupportingInformation}

\setcounter{tocdepth}{1}
\tableofcontents

\section{Experimental methods}
\subsection{Electrochemical cell design and assembly}
\label{sec:exp}
An \textit{in-situ} electrochemical cell optimised for in-plane neutron imaging and diffraction investigations has been developed. A schematic of the custom-made electrochemical cell is presented in Figure~1a in the main text. The main features of this cell design are a cylindrical in-plane electrode alignment and an adjustable electrode height (layer thickness). This enables in-plane imaging and thus depth profile observations. The adjustable electrode height allows the usage of different electrode layer thicknesses.

The electrochemical cell has two electrodes with 16 mm diameter, a thick graphite-based active electrode (400 $\mu$m) -- the region of interest in this study -- and a metallic Li reference electrode (Li ribbon, 99.9~\%, Merck, 750 $\mu$m thick). The electrodes are separated by a glass microfiber filter (Whatman Grade GF/A, 260$\mu$m thick). The three components were immersed in electrolyte  (1M Lithium hexafluorophosphate in Ethylene Carbonate: Dimethyl Carbonate (1M LiPF6 EC/DMC=1:1 vol.\%), BASF) and compressed between current collectors of stainless steel (316SS - an electrochemically stable material). Altogether, these were encased in a neutron transparent Al cell housing with an inner diameter of 18~mm and an isolating inner polytetrafluoroethylene (PTFE) layer with a thickness of 1~mm in the imaged area. The cell was sealed by O-rings, which were fixed between the casing, the polyether ether ketone (PEEK) clamping rings and the current collectors.

The thick graphite-based electrode was prepared in three steps. 
First, 88~wt.\% natural graphite powder (Formula BT, SLC 1520P, Carbo Tech Nordic ApS) as the active material was dry mixed in a mortar with 3~wt.\% of conductive carbon (carbon black, Cabot Corporation, USA), and 9 wt.\% of PTFE (Polytetrafluoroethylene, Sigma-Aldrich) as the binder. Second, the homogeneous powder mixture was pressed into a 16~mm-diameter disk pellet by applying 3 tons of pressure with a SPECAC press. Finally, the pellet was dried in B\"{U}CHI Glass Oven B-585 vacuum furnace within a glovebox with Argon atmosphere. The temperature in the oven was set to 80$^\circ$C for 4~hours and further increased to 120$^\circ$C for 10~hours. The pellet was kept under Argon atmosphere thereafter until the electrochemical cell was assembled, i.e. a day prior to the experiment allowing electrolyte to disperse through the porous electrode.

\subsection{\textit{Operando} neutron investigation details}
\begin{figure}
    \centering
    \includegraphics[scale=0.7, keepaspectratio]{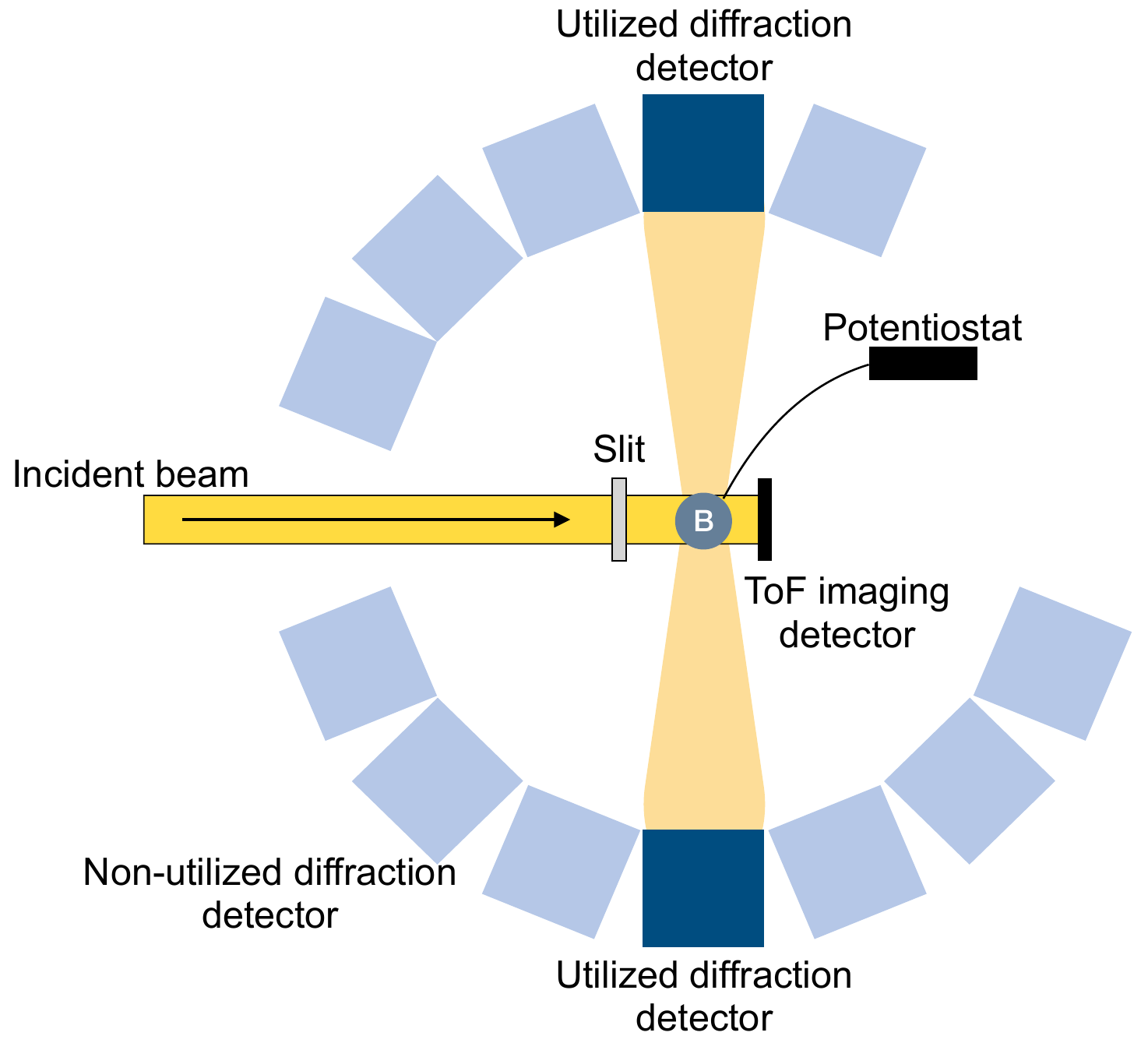}
    \caption{Schematic of utilised neutron instrument set-up based on the SENJU diffractometer at J-PARC, retrofitted with a time-of-flight imaging detector for spatially resolved transmission measurement. B indicates the battery sample.}
    \label{fig:set-up}
\end{figure}

Neutron time-of-flight (ToF) measurements carried out using a polychromatic cold neutron beam, having a wavelength range from 2.0-6.4~\AA, and a Maxwellian spectrum peaking at 2.3~\AA. The neutron flux was approximately $6.5\times10^7$~n~s$^{-1}$cm$^{-2}$ at the sample position at 34.8~m from the source, which operated at a power of 500~kW.

For ToF neutron imaging, a neutron imaging microchannel plate detector \cite{TREMSIN2020} was added to the set-up of SENJU diffractometer. The detector was placed directly behind the sample position in order to utilise the intrinsic detector resolution of 110~$\mu$m. The field-of-view of the detector was 28.16$\times$28.16~mm$^2$ featuring a 512$\times$512 pixel array with a pixel-size of 55~$\mu$m.

Two different time scales have to be considered for a time-resolved ToF imaging experiment, the ToF and the scanning time. The scanning time gives the resolution of the imaged process, which was set by an acquisition time of 15 minutes plus additional tens of seconds for data read-out. Each of these time bins of approximately 15 minutes resolving the process consists of 1774 individual ToF images accumulated with a 25~Hz repetition frequency, corresponding to the neutron source frequency.
The time bin of these 1774 ToF images is 20.48~$\mu$s, which corresponds to a wavelength resolution of $\Delta\lambda/\lambda=0.24$\% for the probed spectrum.
Five read-out gaps, each of 400~$\mu$s, were needed by the detector to read-out the data before it reached saturation. This results in small gaps in the probed spectrum, which can however be distributed strategically to avoid the loss of regions of interest in the spectrum. 

The SENJU diffractometer has 12 detector modules surrounding the sample position at 800~mm (equatorial distance). Each module contains three scintillator area detectors stacked vertically (25~mm vertical gap between detectors), each with an area of 256$\times$256~mm$^2$, divided into 64$\times$64 pixels with the pixel size of 4$\times$4~mm$^2$, offering an angular coverage of approximately 0.103~steradians per detector~\cite{Kawasaki2014}. However, the retrofitted ToF imaging detector downstream of the sample obscured most forward scattering detector banks. Thus a limited detector coverage was utilised and the 90-degree diffraction detector modules (in relation to the incident beam direction) cover a crystallographic d-spacing range of 1.7-4.4~\AA~for the utilised ToF spectrum. Graphite, \ce{LiC12}, and \ce{LiC6} display their most intense diffraction peaks in this range.
These modules provide $\Delta d/d=0.011$ d-spacing resolution at FWHM. During the electrochemical cycle, neutron events were collected continuously. Each neutron event is recorded with an associated detector position and a timestamp.

For the measurements, the beam was limited to 1~mm height in front of the sample, in order to limit scattering from other parts of the sample except the region of interest. Neutron data was acquired \textit{in-situ} during galvanostatic cycling conditions. The \textit{in-situ} as-assembled (pristine) electrochemical cell was operated at a constant current in the voltage range of 0.001–3.000~V~vs.~\ce{Li/Li^+} for one cycle.
The voltage profile of the electrochemical cell is plotted in Figure 2b and 3a in the main text.
The theoretical capacity of graphite was taken as 372~mAh~g$^{-1}$, based on \ce{LiC6} as the fully lithiated state. The cell was cycled at 1.369~mA, corresponding to C/35 cycling rate, as calculated from the theoretical capacity of graphite. 
The cell exhibits approximately 286~mAh~g$^{-1}$ capacity during the first lithiation, excluding the apparent capacity during the SEI formation, and reaches a delithiation capacity of 246.7~mAh~g$^{-1}$.
The cycling was done using a CompactStat electrochemical workstation by Ivium.

\subsection{Diffraction contrast neutron imaging}

While attenuation contrast imaging and neutron diffraction are well-known techniques~\cite{Siegel2011,Zhou2016,Vadlamani2014, TRUCANO1975} for structural investigations on the macroscopic and atomic length scales, respectively, diffraction contrast imaging is a less conventional approach~\cite{WORACEK2018}. 

Diffraction contrast neutron imaging is based on recording wavelength-resolved images and analysing pixel-wise wavelength-dependent features of the transmission spectra. The attenuation of the neutron beam is caused by both absorption and scattering. While the wavelength dependence of the absorption is smooth and moderate in the cold neutron regime, in particular, the coherent elastic scattering of crystalline materials can have a distinct signature on the spectrum due to Bragg diffraction. For a polycrystalline material with random grain orientation, so-called Bragg edges appear in the attenuation spectrum at wavelengths $\lambda=2d_{hkl}$  where the scattering angle $\theta=90^\circ$ and thus no longer wavelengths can be diffracted at the corresponding \textit{hkl} lattice planes. Correspondingly, the transmission is rising steeply beyond such wavelength constituting the Bragg edge. Analyses of these edges, thus, enables the identification of crystallographic phases, lattice strains and signatures of texture, similar to diffraction data, but locally for every pixel of the imaging detector~\cite{WORACEK2018}.

\begin{figure}
    \centering
    \includegraphics[scale=0.46, keepaspectratio]{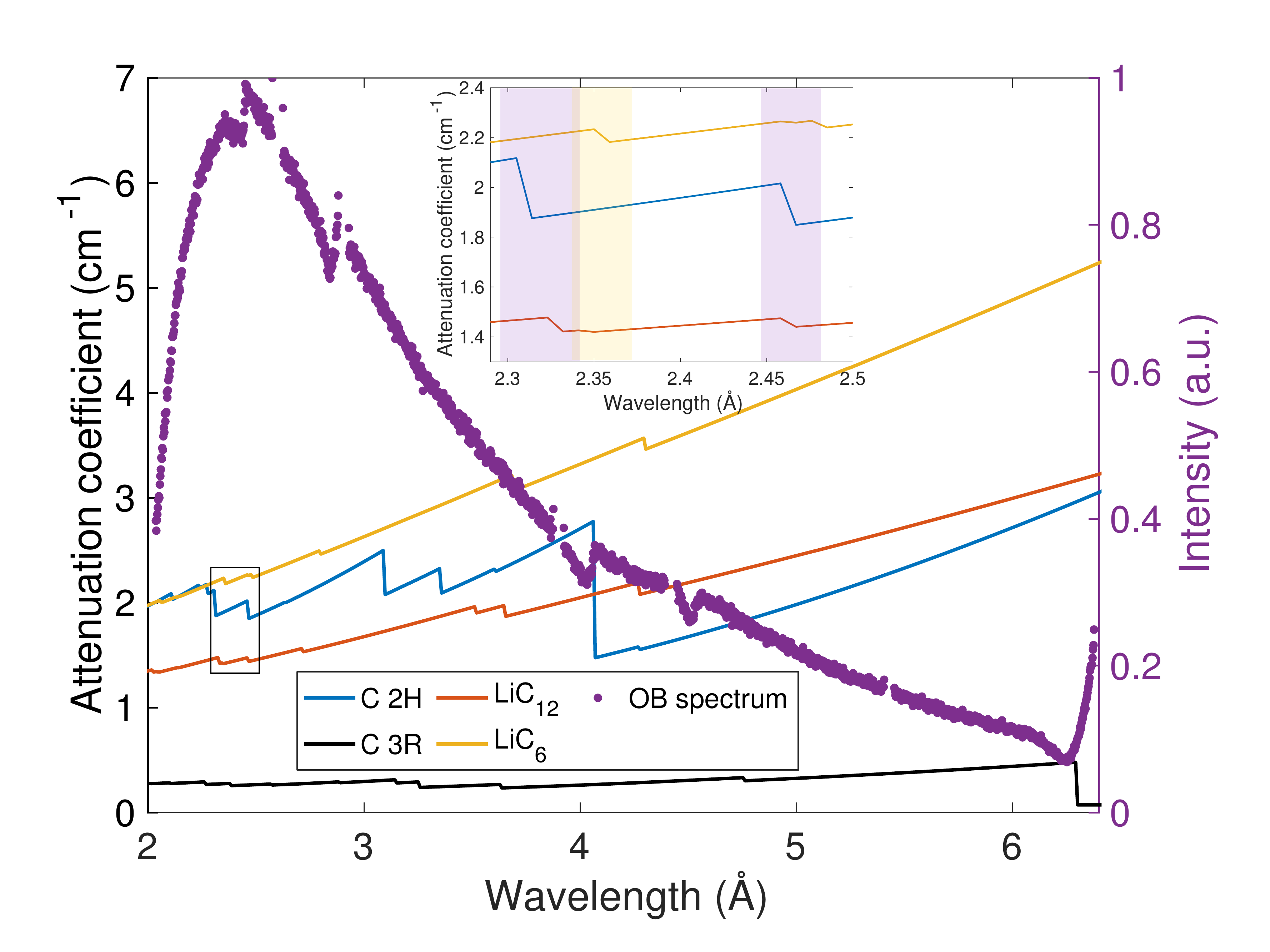}
    \caption{Left y-axis: Attenuation coefficient spectrum of main phases (graphite - hexagonal structure (2H), graphite - rhombohedral structure (3R), \ce{LiC12} and \ce{LiC6}) found in electrochemically lithiated graphite. Right y-axis: Measured neutron spectrum at SENJU. Inset: Zoomed section of the spectrum with the Bragg edges of interest in this study: at  $\lambda=2.31$~\AA~for graphite 2H (1~1~2), at $\lambda=2.33$~\AA~for \ce{LiC12}~(3~0~2), at $\lambda=2.36$~\AA~for \ce{LiC6}~(3~0~1), at $\lambda=2.46$~\AA~for graphite 2H~(1~1~0), and at $\lambda=2.46$~\AA~for \ce{LiC12}~(0~0~3). 
    }
    \label{fig:si:nxs}
\end{figure}

Figure~\ref{fig:si:nxs} displays the theoretical attenuation spectra for the main phases of the graphite electrode during cycling. The graphite~3R spectrum contribution was negligible. The spectra were calculated with the dedicated program nxsPlotter~\cite{Boin2012a}. Figure~\ref{fig:si:nxs} inset focuses on the Bragg edges utilised for the analyses in this work, more specifically the hashed regions. These are (i) the graphite~2H (1~1~2) edge at $\lambda=2.31$~\AA~and the \ce{LiC12} (3~0~2) edge at $\lambda=2.33$~\AA. With commencing lithiation of the graphite, this edge moves gradually from the graphite to the \ce{LiC12} edge position and thus clearly indicates this phase transformation. Analysing the edge position in this regime for each pixel of the recorded ToF image data sets, therefore, provides information on the local lithiation state with respect to this lithiation phase (compare panel (c) in Figure~3 in the main paper).  
(ii) the (3~0~1) Bragg edge of \ce{LiC6}, which is located at $\lambda=2.36$~\AA, is sufficiently distant from other Bragg edges of the system to be analysed independently. Its height will, thus, correlate with the amount of \ce{LiC6} formed in the electrode (compare Figure~3d in the main text), 
(iii) the graphite~2H (1~1~0) Bragg edge is found at $\lambda=2.46$~\AA~and coincides with the \ce{LiC12} (0 0 3) Bragg edge at the same wavelength. Thus this Bragg edge will be present as long as either the graphite phase or the \ce{LiC12} phase will be present at the analysed location of the sample/image, but it will weaken and disappear upon transformation to \ce{LiC6}, which has no significant Bragg edge in this region (Compare panel Figure 3e in the manuscript). The greater height of the carbon edge even enables, to some extent, tracing of the transformation of the graphite phase to the lithiated \ce{LiC12} phase, when assuming that the transformation to \ce{LiC12} is complete throughout the integrated location before the transformation to \ce{LiC6} happens.
As a consequence, the Bragg edge position and height analyses of edges at these three positions enable a complete picture of the local phase transformation kinetics during the cycling process.

\section{Data processing}
\subsection{Image analysis}

Data analysis was focused on a region of interest of $12\times324$ pixels, i.e.  $660~\mu$m~$\times$~17.8~mm. The data was normalised by open beam (i.e. no sample in the beam) images to obtain transmission images. The transmission values were transformed to projected attenuation coefficient values based on the known sample thickness and the Beer-Lambert law. Due to the evident lateral uniformity, the data was averaged horizontally (along x-direction), which improved statistics and enabled the intuitive visualisation, by utilising correlated panels including time dependence and depth profile resolution of the graphite electrode. 

The corresponding wavelength-resolved imaging data of 12 spectra per process time bin had to be further analysed to be presented in several panels. ToF was converted to wavelength according to the de Broglie relationship. The regions of the selected Bragg edges, 100 wavelength bins centred on the nominal position of the edges (Figure~\ref{fig:si:nxs}) were selected. A Fourier filtering function was applied to the segments of data points to smooth the data. The local maximum and minimum before and after the nominal Bragg edge position were determined, respectively. Subsequently, three linear functions separated by the maximum and minimum points encompassing the edge were fitted according to the procedure of Ref.~\cite{Carminati2020}. The fit determined the position, height, and width of the edges. The three fitted parameters were checked for validity, and the pixels where fitting failed, were disregarded (compare Figure 3c-e in the manuscript). For the three Bragg edge regions of interest, the respective required fitting parameters were extracted for each pixel and plotted into the three diffraction contrast panels presented in Figure 3.

\subsection{Powder diffraction analysis}

For the upper 90-degree detectors, the raw data was reduced to 1-dimensional histograms, by radially integrating over the scattering coverage with 0.002~\AA~d-space binning. The data from both opposing detectors were combined in the 3.2-3.7~\AA~d-range for improved statistics. The data was reduced to time bins matching the time resolution of the imaging data, i.e. approximately 15 min. 
The peaks were individually fitted as follows: the intense single peaks were fitted with an asymmetric pseudo-Voigt function, the overlapping double peaks were fitted with a combination of pseudo-Voigt functions, and low-intensity peaks from \ce{LiC6} at 3.699~\AA~were fitted by a Gaussian function.

\section{Additional data and figures}

\begin{figure}
    \centering
    \includegraphics[scale=0.48, keepaspectratio]{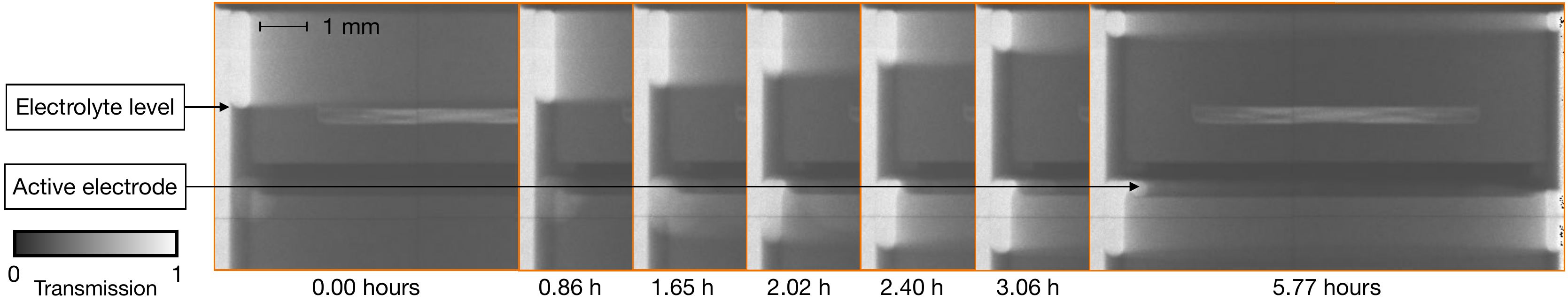}
    \caption{Attenuation contrast images of cell during lithiation, overlapping in a time series, to illustrate a significant electrolyte movement away from the active electrode with and against gravity.}
    \label{fig:sup:electrolyte}
\end{figure}

In the main paper, we discuss the influence and bias of electrolyte on the observation of Li redistribution during lithiation-delithiation processes. Here, we illustrate these effects by two observations related to pre-studies of the presented measurements of the main paper. Figure~\ref{fig:sup:electrolyte} shows a time sequence of attenuation contrast images of a similar cell recorded at the neutron imaging instrument RADEN at J-PARC~\cite{Shinohara2016}. The images are taken during lithiation, while the cell was lithiated at a C/34 rate. The sequence illustrates the significant redistribution of electrolyte due to its polarisation in a cell only partially filled with electrolyte. In the incipient state at OCV, the electrolyte, affected by gravity, fills evenly the lower part of the cell surrounding the electrochemically active electrode. When current is applied over the course of several hours, parts of the electrolyte migrate against the ionic flow and against gravity to regions above the separator, leaving the graphite electrode covered with an uneven thin film. The images also illustrate the strong effect of the electrolyte filling on the contrast achieved and the potential for detailed observations of the electrolyte distribution and redistribution in a battery by neutron imaging.

\begin{figure}
    \centering
    \includegraphics[scale=0.7, keepaspectratio]{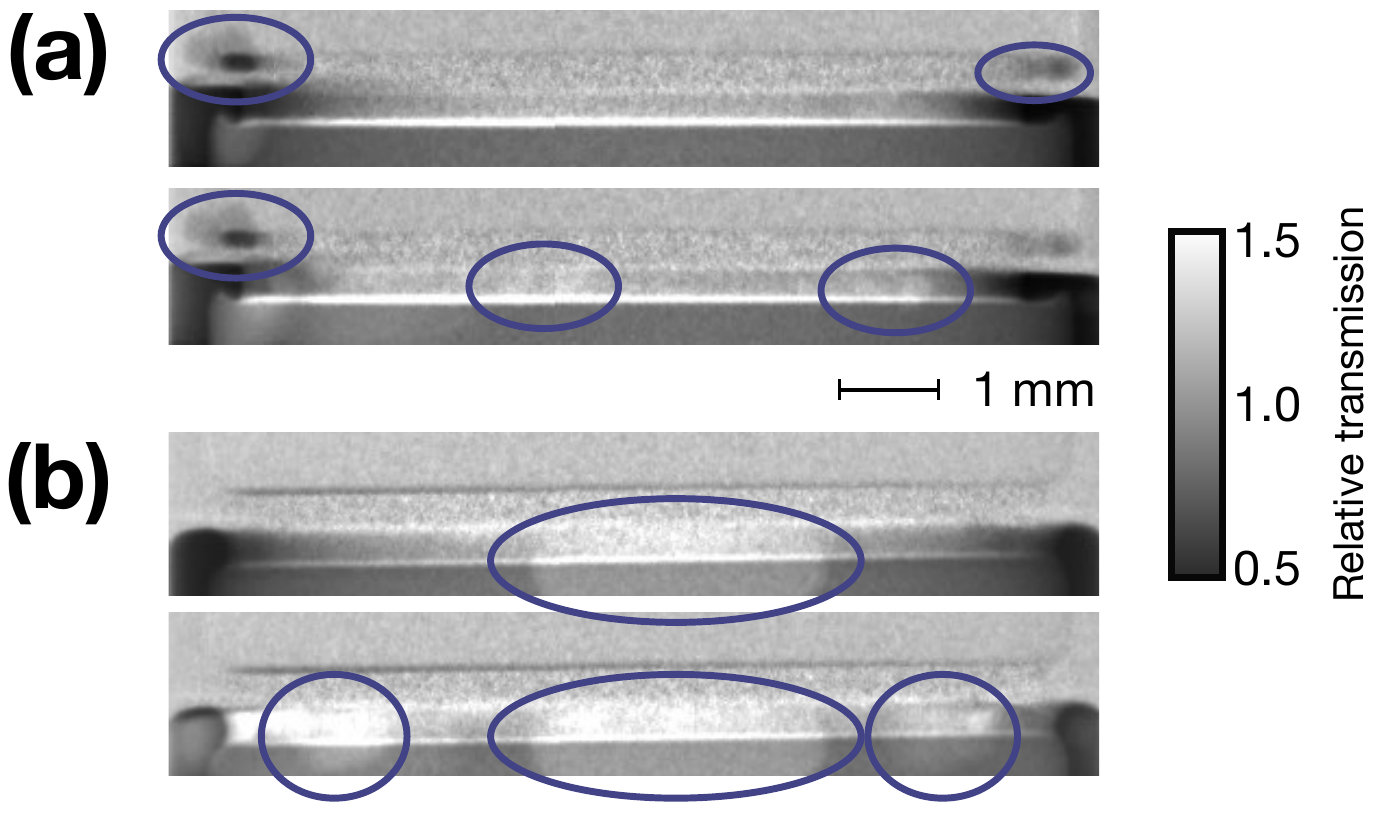}
    \caption{Detail images of relative attenuation contrast of two (a and b) cells (each normalised to its lithiated state) illustrating the impact of irregular changes of electrolyte distribution due to e.g. bubbles and reservoirs (highlighted by blues circles).}
    \label{fig:sup:gas}
\end{figure}

However, in particular when the electrolyte distribution is irregular, as e.g. in the case of bubbling and small reservoirs with changing filling state, the electrolyte contribution is difficult to disentangle from the Li redistribution in attenuation contrast images, and thus there is a significant risk of bias to the analyses of Li and electrolyte redistribution. Figure~\ref{fig:sup:gas} illustrates situations, where such irregular distributions become obvious. Figure~\ref{fig:sup:gas} shows examples from two samples at different charging states, which, for better visibility, have been normalised to the fully lithiated state of the respective sample. Thus, bright areas are created by lower attenuation as compared to the lithiated state, while darker areas indicate more attenuation. Some distinct features that cannot be attributed to regular Li exchange but resembling bubbles and some different partial filling reservoirs are clearly identifiable in this case.

\begin{figure}
    \centering
    \includegraphics[scale=0.70, keepaspectratio]{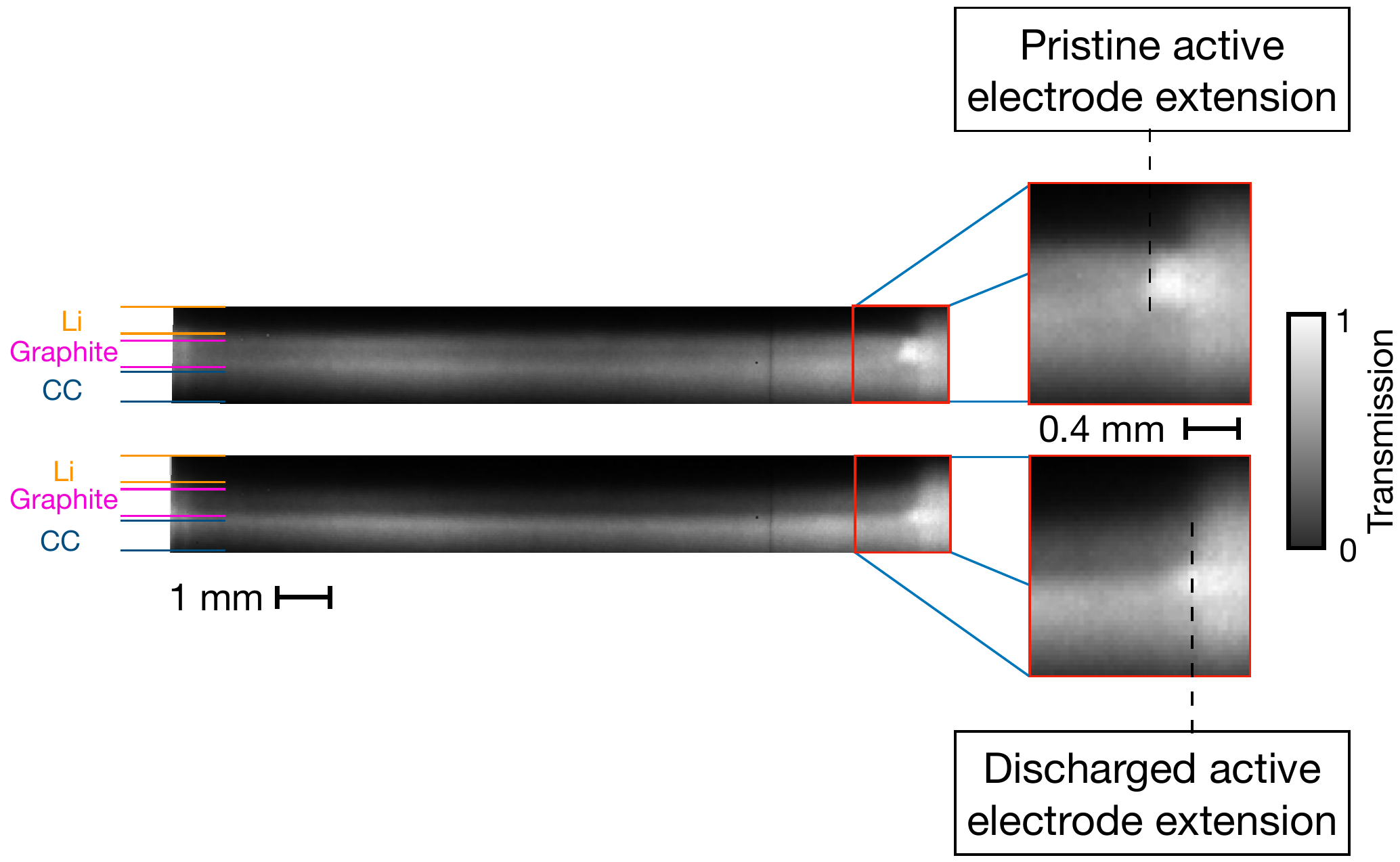}
    \caption{Transmission images showing the horizontal expansion of the active electrode. The corresponding components are indicated.}
\label{fig:sup:expansion}
\end{figure}

Another macroscopic effect observed with attenuation contrast imaging is the volumetric expansion of the graphite electrode. Figure~\ref{fig:sup:expansion} shows the horizontal expansion of the active electrode upon lithiation.

\begin{figure}
    \centering
    \includegraphics[scale=0.46, keepaspectratio]{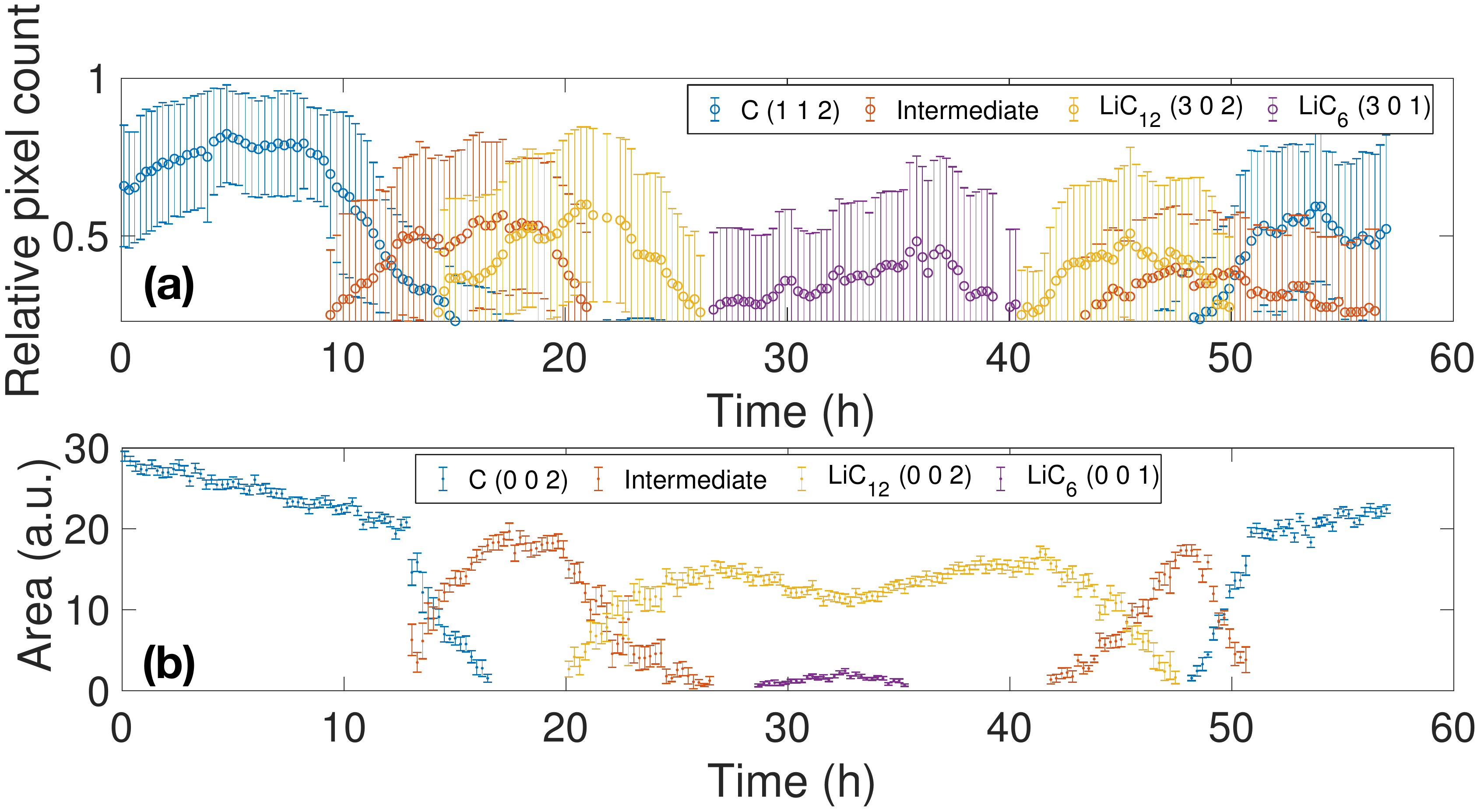}
    \caption{Diffraction contrast pixel count relative to different identified phases (a) versus analysed diffraction peak area (b) plotted against process time.}
\label{fig:sup:counts}
\end{figure}

Finally, as stated in the manuscript Figure~\ref{fig:sup:counts} illustrates an integral correlation of the diffraction contrast imaging data and the diffraction results. A qualitative resemblance can be achieved despite the high noise level of the imaging data, where for this comparison a simple pixel count after application of a threshold for the Bragg edge position displayed in Figure 3c and d has been utilised for the evaluation of the graphite, intermediate and \ce{LiC12} phase. The \ce{LiC12} phase disappears in the imaging data due to a vanishing signal in this range at the high attenuation levels implied by the high lithiation and \ce{LiC6} phase, which is evaluated here from the panel (d) in Figure 3.

\newpage
\printbibliography